\documentclass[aps,pre,reprint,showpacs,groupedaddress]{revtex4-1}
\usepackage{graphicx}
\usepackage{hyperref}
\usepackage{amsmath}
\usepackage[usenames, dvipsnames]{color}
\usepackage{lineno}
\hypersetup{pdfborder={0 0 0}}

\begin{document}
\title{Pseudo-spectral Maxwell solvers for an accurate modeling of Doppler harmonic generation on plasma mirrors with Particle-In-Cell codes}
\date{Submitted on $23^{th}$ August 2016}
\author{G. Blaclard$^{1}$}
\author{H. Vincenti$^{1,2}$}
\author{R. Lehe$^{1}$}
\author{J.L. Vay$^{1}$}
\affiliation{$^{1}$Lawrence Berkeley National Laboratory, 1 Cyclotron road, Berkeley, California, USA}
\affiliation{$^{2}$Lasers Interactions and Dynamics Laboratory,  CEA Saclay, France}

\begin{abstract}
With the advent of PetaWatt (PW) class lasers, the very large laser intensities attainable on-target should enable the production of intense high-order Doppler harmonics from relativistic laser-plasma mirrors interactions. At present, the modeling of these harmonics with Particle-In-Cell (PIC) codes is extremely challenging as it implies an accurate description of tens to hundreds of harmonic orders on a a broad range of angles.  In particular, we show here that due to the numerical dispersion of waves they induce in vacuum, standard Finite Difference Time Domain (FDTD) Maxwell solvers employed in most PIC codes can induce a spurious angular deviation of harmonic beams potentially degrading simulation results. This effect was extensively studied and a simple toy-model based on Snell-Descartes law was developed that allows us to finely predict the angular deviation of harmonics depending on the spatio-temporal resolution and the Maxwell solver used in the simulations.  Our model demonstrates that the mitigation of this numerical artifact with FDTD solvers mandates very high spatio-temporal resolution preventing doing realistic 3D simulations even on the largest computers available at the time of writing. We finally show that non-dispersive pseudo-spectral analytical time domain solvers can considerably reduce the spatio-temporal resolution required to mitigate this spurious deviation and should enable in the near future 3D accurate modeling on supercomputers in a realistic time-to-solution. 
\end{abstract}

\maketitle

\section{Introduction}

\subsection{Scientific context: Doppler harmonic generation on plasma mirrors }

The advent of high power PW femtosecond lasers \cite{STRICKLAND1985219} opened the way to a new, promising but still largely unexplored branch of physics called Ultra-High Intensity physics (UHI). When such a laser is focused on a solid target, the laser intensity can reach values as large as $10^{23}W.cm^{-2}$, for which matter is fully ionized and turns into a plasma mirror that reflects the incident light \cite{Kapteyn:91,PhysRevE.69.026402,thaury2007}. The corresponding laser electric field at focus is so high that plasma mirror particles (electrons and ions) get accelerated to relativistic velocities upon reflection of the laser on its surface \cite{RevModPhys.78.309}. A whole range of compact tabletop sources of high-energy particles (electrons, protons, highly charged ions) and radiations ranging from UV to $X$-rays may thus be produced from the interaction between this plasma mirror and the ultra-intense laser field at focus. 

In this study, we will focus on the case of high-order Doppler harmonics generated on relativistic plasma mirrors  \cite{Dromey2006,PhysRevLett.99.085001,thaury2007} . For intensities such as the ones achieved by 100TW/PW lasers, the very high laser electric field drives periodic oscillations of the plasma mirror surface at relativistic velocities. These periodic oscillations induce a periodic distortion of the reflected field by Doppler effect  \cite{Thaury2010,:/content/aip/journal/pop/3/9/10.1063/1.871619,PhysRevE.74.046404,PhysRevE.84.046403}, which is associated in the frequency domain to a comb spectrum made of high harmonics of the incident laser frequency.

There are two main motivations for understanding and controlling the spatio-temporal properties of this radiation source. The first motivation is purely fundamental. A lot of physical information from the laser-plasma interaction is "encoded" in the harmonic spectrum which can then be used  as a probe of the relativistic plasma mirror dynamics \cite{PhysRevLett.108.113904,Wheeler2012,Vincenti2014}. The second motivation is to use this radiation source for performing innovative time-resolved application experiments. By filtering a group of these high-order harmonics, one can generate intense attosecond pulses of light \cite{1367-2630-8-1-019,PhysRevLett.108.113904,Wheeler2012} in the time domain. Those are currently considered as one of the best candidate light sources for the very first attosecond pump-attosecond probe experiments to demonstrate the ultimate goal of "filming" electron dynamics in matter. 

All these promising perspectives have attracted a lot of interest from the laser-plasma community and an X-ray beam based on plasma mirror harmonics is therefore scheduled at the future Extreme Light Infrastructure (ELI) in Europe.

\subsection{Main challenges in the numerical modeling of Doppler harmonic generation}

The success of the ELI and future attosecond beamlines based on this type of radiation source will rely on the strong coupling between experiments and large-scale simulations with Particle-In-Cell (PIC) codes. 

The modeling of high-order Doppler harmonic generation with PIC codes is however very challenging as it involves an accurate description of a large band of frequencies sometimes spanning wide angles in the relativistic regime (due to the curvature of the plasma mirror that acts as a focusing mirror which naturally increases the harmonic beam divergence) \cite{Vincenti2014}. A realistic modeling of harmonic generation therefore requires Maxwelll solvers that do not induce any numerical dispersion of all harmonic orders in a rather large range of angles of emission. 

We show here that standard PIC codes currently in use employing Finite-Difference Time-Domain Maxwell solvers (FDTD) to advance electromagnetic fields in time and space induce a non-negligible numerical dispersion of electromagnetic waves that can significantly degrade harmonic spectra. Indeed, due to the numerical dispersion of these Maxwell solvers, vacuum artificially acts as a dispersive medium that can deviate harmonic beams when these enter vacuum at the vacuum-plasma interface. In the worst case, this effect could even lead to a wrong interpretation of experimental results with PIC codes.   

To address this challenge and enable realistic modeling of Doppler harmonics, one solution is to use highly precise and dispersion-free pseudo-spectral methods to solve Maxwell's equations.  Despite their accuracy, such methods have however hardly been used during the last two decades in PIC codes due to their low scalability to $10,000$s of cores at best, which is not enough to take advantage of petascale supercomputers architectures required for 3D modeling. To break this barrier a pioneering grid decomposition technique \cite{Vay2013260} was recently proposed for very high-order/pseudo-spectral solvers that was first validated by an extensive analytical work \cite{Vincenti2016147} and then implemented and benchmarked in our high performance WARP+PXR PIC code \cite{Warp, 2016arXiv160102056V}. 

The accuracy brought by pseudo-spectral methods coupled by this highly efficient parallel implementation in WARP+PXR could enable for the first time realistic 3D simulations of Doppler harmonic generation on large petascale machines

\begin{table*}
\begin{tabular}{ccccc}
Class solver & Solver & \begin{tabular}{c} Discretization \\ space and time \end{tabular} & \begin{tabular}{c} Courant condition \\ ($\eta=c\Delta t/\Delta x$) \end{tabular} &\begin{tabular}{c} Refractive index \\ $n=c/v_{\varphi}$ \end{tabular}\\
&&& \\ 
\hline\hline &&& \\ 
 & Yee & \{ $\nabla_2, \delta_t$ \} &$\eta \leq 1/\sqrt{2} $ &$ \frac{|\mathbf{k}|c \Delta t}{2}. 1/arcsin \left( \frac{c\Delta t}{2}. \sqrt{\frac{S_x^2}{\Delta x^2}+\frac{S_z^2}{\Delta z^2}} \; \right) $\\ &&& \\ 
 \begin{tabular}{c} Finite Difference \\ Time Domain \end{tabular} &  \begin{tabular}{c} Cole-Karkkainen-Cowan \\ (CKC) \end{tabular}  & \{ $\nabla_2^\beta, \delta_t$ \} &$\eta \leq 1$ & $ \frac{|\mathbf{k}| c \Delta t}{2}. 1/arcsin  \left( \frac{c\Delta t}{2}. \sqrt{A_x\frac{S_x^2}{\Delta x^2}+A_z\frac{S_z^2}{\Delta z^2}} \; \right)$ \\ &&& \\ 
 & FDTD p-order & \{ $\nabla_p, \delta_t$ \} &$\eta \leq \left(\sqrt{2} \; \sum_{l=1}^{p/2} \left\vert C_l^p \right\vert \right)^{-1}$ & $\frac{|\mathbf{k}| c \delta t}{2}. 1/arcsin  \left(c\Delta t.\sqrt{ \Sigma^2_{p,x} + \Sigma^2_{p,z} } \; \right)$  \\ &&& \\ 
\hline 
&&& \\ 
& PSTD & \{ $k, \delta_t$ \} &$\eta \leq \sqrt{2}/\pi$ & $\frac{|\mathbf{k}| c \Delta t}{2}. 1/arcsin  \left( \frac{c \Delta t}{2}\sqrt{k_x^2+kz^2} \; \right)$ \\ &&& \\ 
\begin{tabular}{c} Pseudo-Spectral \\ Time Domain \end{tabular} & PSATD & \{ $k, \partial_t$ \} &None & $1$ \\&&& \\ 
& PSATD p-order & \{ $k_p, \partial_t$ \} &None & $\frac{|\mathbf{k}| c \Delta t}{2}. 1/ \sqrt{ \Sigma^2_{p,x} + \Sigma^2_{p,z} } $ \\ &&& \\ 
\hline
\end{tabular} 
\caption{Refractive index equation and Courant condition for each scheme in 2D. We use $S_X=sin \left( \frac{k_X \Delta X}{2} \right)$ and $A_X=(1-\beta)+\frac{\beta}{2}\cos(k_Y\Delta Y)$ with $\beta$ a coefficient defined in \cite{Vay20115908}.  For p-order solver, $\Sigma_{p,X} = \sum_{l=1}^{p/2} C_l^p \frac{S_{l,X}}{\Delta X}$ where $S_{l,X}=sin \left( \frac{(2l-1)k_X \Delta X}{2} \right)$ and $C^p_l$ are Fonberg coefficients defined in \cite{Fornberg1990} but for a staggered grid as explained in \cite{Vincenti2016147}. In the Discretization column, $\delta_t$ and $\partial_t$ mean respectively the finite discretization and the analytical resolution in time. In space, $\nabla_p$ is the p-order finite discretization of spatial derivatives, corresponding to a p-order stencil. $k_p$ is the Fourier transform $\hat{\nabla}_p$ of $\nabla_p$. When p $\rightarrow \infty$, $k_p$ tends to $k$, analytical solution in Fourier space.  Moreover, note that only PSATD and CKC solver present a Courant condition that allows $\Delta t=\Delta x$.}
\label{Table_n}
\end{table*}

\subsection{Goals and outline of this study}

In this paper, we extensively study the effect of the numerical dispersion induced by different Maxwell solvers on the numerical modeling of Doppler harmonic properties with the PIC code WARP+PXR. We show that FDTD solvers induce a non-physical angular dispersion of high-order harmonics that can spoil the simulation results  even at very high resolution. The numerical dispersion induced by these Maxwell solvers modify the effective refractive index of vacuum, which can act as a diopter and angularly disperse frequencies. Thanks to the highly scalable parallel implementation of pseudo-spectral solvers in WARP+PXR, we then show that the use of these Maxwell solvers can completely solve the dispersion issue even at moderate resolution. 

Along this line, the study will be divided in four parts: 

\begin{itemize}
\item In part II, we present the numerical configuration used to simulate Doppler harmonic generation with PIC codes. We then quickly introduce the diagnostics used to study the generated harmonic spectrum and show that the numerical dispersion of waves induced by FDTD solvers can create a non-physical angular dispersion of high-order harmonics. 
\item In part III, we briefly review the numerical dispersion of electromagnetic waves introduced by the most standards second order finite-difference Maxwell solvers used by our community. We then generalize the discussion to solvers at arbitrary order $p$ and in the infinite limit to the pseudo-spectral solvers. This part will be crucial to understand the numerical effect of these solvers on the harmonic spectra and develop the simple toy-model presented in the next section. 
\item In part IV, we show that the non-physical angular dispersion of Doppler harmonics can be interpreted as a simple refraction of harmonics by the plasma-vacuum interface. We develop a simple refraction model and validate it against simulation results obtained with various solvers. 
\item In part V, we show the benefits of dispersion-free pseudo-spectral solvers against FDTD solvers. We then estimate a potential speed-up gain in using these solvers both in terms of time-to-solution and memory. 
\end{itemize}

\section{Numerical simulation of Doppler harmonic generation on plasma mirror}

In this section we first introduce the 3D PIC code we used in this study to perform numerical modeling of Doppler harmonic generation. We then briefly introduce a typical simulation configuration for Doppler harmonic generation and present the diagnostics used to compute harmonic spectra in the far field. Finally, we illustrate the non-physical features induced by the Maxwell solver on harmonic spectra that need to be characterized. 

\subsection{The WARP+PXR Particle-In-Cell code}

In this study, we used the WARP+PXR 3D PIC code that we briefly review in this section. 

\subsubsection{The WARP legacy code}
WARP \cite{Warp_url} is an extensively developed open-source 3D PIC code designed to simulate a rich variety of physical processes including laser-plasma interactions at high laser intensities. WARP is written in a combination of 1) Fortran for efficient implementation of computationally intensive tasks 2) Python for high level specification and control of simulations and 3) C for interfaces between Fortran and Python. WARP has now been routinely used for many years on NERSC supercomputers (MCurie/Seaborg/Bassi/Franklin/Hopper/Edison/Cori) and other platforms by many scientists worldwide.

 \begin{figure*}
\includegraphics[width=0.8\linewidth]{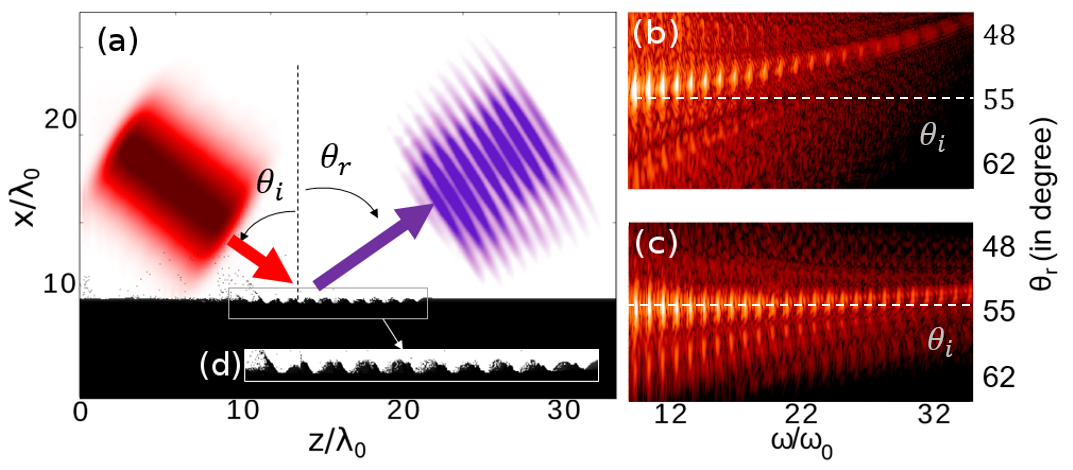} 
\caption{Simulation configuration and effects of FDTD Maxwell solvers on harmonic spectra. For all simulations presented on this figure, an incident laser (p-polarized), normalized amplitude $a_0=8$, waist of $6 \; \lambda_0$, gaussian spatial profile, duration $\tau_0=10 \; T_0$ with $T_0$ the laser period, hypergaussian temporal profile) impinges with an angle $\theta_i=55$ degrees on the plasma mirror. The plasma mirror has an initial exponential density profile of gradient scale length $L=\lambda_0/20$ where $\lambda_0$ is the laser wavelength, equal to $0.8 \; \mu m$. The simulation box dimensions are $40 \lambda_0\times40\lambda_0$ in (x,z) directions. The FDTD order 2 Cole-Karkkainen-Cowan (CKC)  Maxwell solver was used. (a) Spatial evolution of $B_y$ field and electron density at different time. In this simulation the spatial resolution was $\Delta x=\Delta z=\lambda_0/70$ in (x,z) directions. $25$ plasma particles per cells were used. The spatial laser amplitude profile ($B_y$ component) is sketched in red at a a time $t=24.6 \; T_0$ before reflection. The reflected field spatial amplitude profile ($B_y$ component filtered between harmonic orders $4$ to $8$) is represented in violet at a time $t=52.7 \; T_0$ after reflection of the laser on the plasma mirror surface. The plasma mirror electron density spatial profile is represented in black at a time $t=39.13 \; T_0$ during the laser-plasma mirror interaction. Panel (d) shows zoom of the plasma mirror surface. (b) Angularly resolved spectrum (logscale) of the reflected field computed from the same simulation as in (a) i.e with a spatial resolution of $\Delta x=\Delta z=\lambda_0/70$ in (x,z) directions. (c)  Angularly resolved spectrum (logscale) of the reflected field computed from the same simulation as in (a) but with a higher spatial resolution of $\Delta x=\Delta z=\lambda_0/140$ in (x,z) directions. In all cases, the time resolution used is fixed by the Courant condition of the CKC scheme  (cf Table \ref{Table_n}).}
\label{exp}
\end{figure*}

\subsubsection{The PXR library}
 
 As part of the NERSC \cite{NERSC} Exascale Science Application Program (NESAP \cite{NESAP_url}) a full Fortran 90 high performance PIC library PICSAR (for Particle-In-Cell Scalable Application Ressource - abbreviated in PXR) was recently developed by our team to help porting PIC codes to future manycore-based exascale machines \cite{PXR_url}. PXR can be run as a standalone 3D PIC code or as a library to help boost performances of other PIC codes. It has been recently coupled back to WARP through a python layer, by defining a python class that re-defines most of the time consuming WARP methods of the PIC loop. We nicknamed  this code WARP+PXR. 
 
 The PXR library includes numerous optimization strategies to fully benefit from the three levels of parallelisms (Internode, Intranode, Vectorization) offered by current and upcoming architectures (exascale). In particular, thanks to the developments made in PXR (some developments are detailed in \citep{2016arXiv160102056V}), WARP+PXR is now a highly optimized code and includes MPI dynamic load balancing at the internode level, optimized MPI stencil communications, hybrid MPI/OpenMP parallelization of the PIC loop, particle tiling and sorting for optimal cache reuse/memory locality and good shared memory OpenMP scaling/intra-node load-balancing, threaded FFTWs for the advanced Maxwell solvers, as well as cutting edge SIMD algorithms for efficient vectorization of hotspots routines. Other optimizations notably include use of MPI-IO for efficient parallel dumping of particles and fields. PXR has now been entirely ported to the new Intel KNL architectures and shows very good performances in the early benchmarks done in preparation for NERSC's Cori phase 2. The WARP+PXR simulation tool is now routinely used on NERSC supercomputers in support of laser-plasma experiments performed at CEA Saclay in France on the 100 TW laser UHI100 and also for upcoming laser-plasma experiments planned at LBNL on the BELLA PW laser. The PICSAR library will be released with an open source license in the near future. 
 
 \subsubsection{Highly scalable pseudo-spectral Maxwell solvers in PXR}
 
In PXR, both FDTD and Pseudo-Spectral (PS) solvers, which will be detailed in section III and presented in Table \ref{Table_n}, have been implemented and optimized. In particular, a pioneering implementation of highly scalable pseudo-spectral Maxwell solvers was achieved in PXR and tested at large scale on the MIRA supercomputer at Argonne National Laboratory. Below, we briefly present the principle of this implementation. It will be presented in greater details in a separate paper. 

Our implementation uses the standard domain decomposition used in standard FDTD solvers, where the simulation domain is split in several subdomains with guard regions containing copies of adjacent subdomains. Maxwell's equations are solved in the spectral domain on each subdomain by performing local FFTs instead of harder to scale global FFTs, as in regular pseudo-spectral solvers. This technique is highly scalable and demonstrated very good scaling on up to 800,000 cores on MIRA.  

This technique however implies eventually a small truncation error that was characterized in a recent study \cite{Vincenti2016147}.  This study shows that this error depends on the stencil order $p$, the width of guard regions and the spatio-temporal resolution used in the simulations. In particular, it shows that the best solution to minimize truncation errors is to use very high-order-p Maxwell solvers with a moderately low number of guard cells. This yields the same solution (to machine precision) than infinite order $p\rightarrow \infty$ pseudo-spectral solvers on a large band of frequencies with truncation errors lower than machine precision. 

\subsection{Doppler harmonic simulation case used in WARP+PXR}

\subsubsection{Case presentation and diagnostics used}

As detailed in the introduction, the modeling of Doppler harmonic generation is challenging and usually requires a high spatio-temporal resolution to resolve the high harmonic orders in space and time. In experiments, it is common to observe beyond the 40-50$^{th}$ harmonic orders.  This already mandates a resolution at least lower than $\lambda_0/100$ (Shannon criterion). In addition, the FDTD Maxwell solver used in the standard PIC method induces numerical dispersion of electromagnetic waves that even requires a much higher resolution. This numerical dispersion can produce non-physical degradation of harmonic spectra.

On Fig. \ref{exp}, we present the configuration and the numerical parameters used in a typical 2D-PIC simulation of Doppler harmonic generation on plasma mirrors (panel (a)). The p-polarized incident laser (amplitude profile in red), impinges at oblique incidence with an angle $\theta_i=55$ degrees on the plasma mirror (electron density in black, a zoom on the plasma mirror surface (d)).This angle $\theta_i$ is chosen in order to maximise the harmonic intensity, as explained in \citep{Thaury2010}. The plasma mirror has an exponential density profile with a gradient scale length $L=\lambda_0/20$ and a maximum density of $n_{max}=240n_c$ (plastic target), where $n_c$ is the plasma critical density associated to the laser wavelength.   For the Maxwell solver, a Cole-Karkkainen-Cowan FDTD solver (CKC) \cite{Cole1997,Cole2002,Karkkainen2006,Cowan201161} was used. 

To compute the angularly resolved harmonic spectrum in the far field, we used Fraunhofer diffraction integral on the reflected field given by the PIC code. Angularly resolved spectra are plotted on panels (b) and (c) for two different spatial resolutions  $\Delta x=\Delta z=\lambda_0/70$ and $\Delta x=\Delta z=\lambda_0/140$. The time step is chosen so that $c\Delta t=\Delta x$, maximum authorized value for the CKC solver given in Table \ref{Table_n}.

Note that as the laser impinges on a pre-plasma of gradient scale length $L$, it actually reflects on the plasma mirror at a much lower density than $n_{max}$ so that the skin-depth of the laser is always well resolved in all the simulations presented in this paper and even for the poorly resolved cases.  

\subsubsection{Effect of numerical dispersion on harmonic spectra}

Fig.  \ref{exp} (a) and (d) show that periodic oscillations of the plasma mirror surface are driven by the incident laser.  These oscillations periodically distort the reflected field by Doppler effect and a train of attosecond pulses (amplitude profile in violet) is emitted with an angle $\theta_r$ from the normal to the plasma mirror surface. This attosecond pulse train is associated to a high harmonic comb spectrum in the frequency domain (see angularly resolved spectrum in panels (b) and (c)). 

In principle, Doppler harmonics should be all emitted around the specular direction ($\theta_r=\theta_i$) \cite{Thaury2010,Vincenti2014}. However, we can observe that in the particular case of panel (b) there is an angular deviation of high-order harmonics with respect to the specular direction $\theta_r=\theta_i$: the higher the harmonic order, the larger the angular deviation. 

Panel (c) further shows that this deviation is significantly reduced when the spatio-temporal resolution is increased. The dependence on harmonic order and spatial resolution of this non-physical angular deviation suggests a strong influence of the numerical dispersion of the Maxwell solver on the spatio-temporal properties of Doppler harmonics that explicitly needs to be characterized. 

\section{Influence of the Maxwell solver on the vacuum refractive index}
\label{Influence}

In this section, we analyze the  numerical dispersion of different Maxwell solvers of the PIC algorithm and their influence on the modification of the vacuum refractive index. First, we give a brief overview of the numerical dispersion of Finite Difference Time Difference (FDTD) solvers of  arbitrary order-$p$. Then we analyze the numerical dispersion of Pseudo-Spectral solvers in the infinite order limit $p\rightarrow \infty$.  

\begin{figure}[h!]
\includegraphics[width=\linewidth]{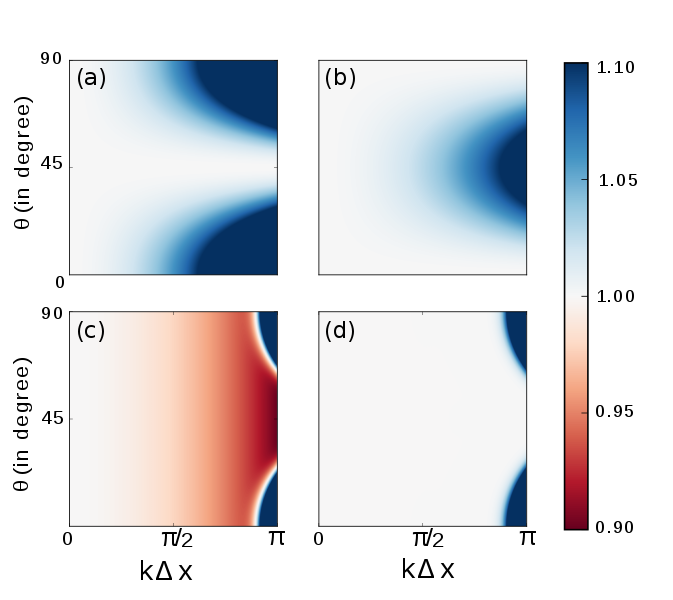} 
\caption{Refractive index $n_r$ of vacuum for different Maxwell solvers as a function of normalized wave numbers $k\Delta x$ and propagation angle $\theta = arctan(k_z/k_x)$ of the electromagnetic waves. For each solver, the time step is given by the Courant condition (cf. Table \ref{Table_n}). (a) Index $n_r$ of vacuum for the Yee solver (b) Index $n_r$ of vacuum for the CKC solver (c) Index $n_r$ of the PSTD-order 128 solver (d) Index $n_r$ of the PSATD-order 128.}
\label{optical_index}
\end{figure}

\subsection{Most common order $p=2$ FDTD solvers}

\begin{figure*}[t!]
\includegraphics[width= \linewidth]{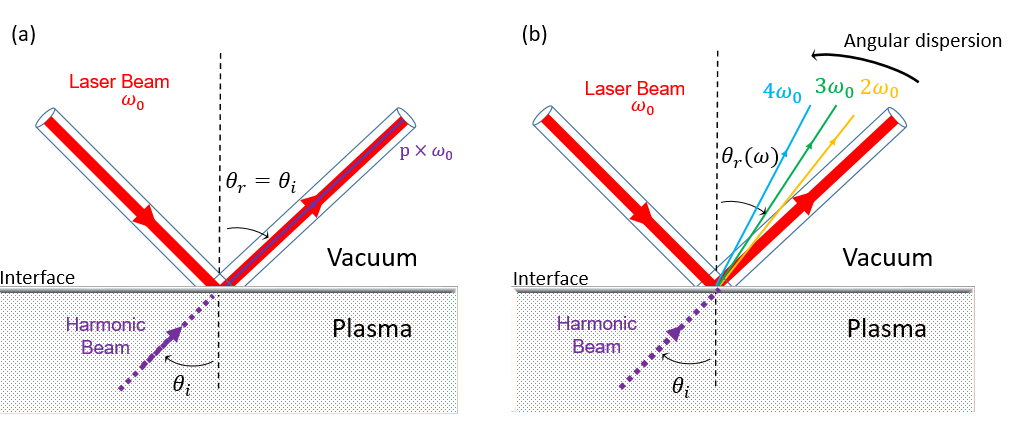} 
\caption{Model for angular deviation of harmonics. The harmonic beam is modeled as an incident beam generated at the plasma-vacuum interface and entering vacuum with an incidence angle $\theta_i$ (a) The physical case is represented. Because the vacuum index is $1$, there is no dispersion and the harmonics are generated at the angle of specular reflection. (b) : the transmission medium is numerical, therefore dispersive. Each frequency is deflected with an angle $\theta_r$, which can be calculated with our simple model based on Snell's law.}
\label{target}
\end{figure*}

In standard PIC codes, Maxwell equations are discretized in space and time using order $p=2$ FDTD solvers.  These solvers have been used mostly for their efficient parallelization on up to millions of cores. However, these solvers introduce non-physical numerical dispersion of electromagnetic waves in vacuum that can be highly detrimental to the numerical modeling of high harmonic generation on plasma mirrors. 

Indeed in vacuum, an electromagnetic wave of frequency $\omega$ and wavevector $k$ should obey the dispersion relation $\omega=kc$, which results in a phase velocity equal to $c$. FDTD solvers modify this dispersion relation and introduce a dependency of the phase velocity on wavevector and frequency. \\
For each solver presented in this paper, the refractive index defined by $n=c/v_\varphi$ is shown in Table \ref{Table_n}. Whereas "real" vacuum refractive index is equal to 1, the "numerical vacuum"  becomes a dispersive medium in PIC simulations. In particular, we show later the plasma-mirror-vacuum interface can act as a refraction boundary that disperses the generated high harmonics. \\

On Fig.\ref{optical_index}, 2D maps of refractive indices are shown for two different order p=2 solvers: the Yee solver (panel (a)) and the Cole-Karkkainen-Cowan (CKC) solver (panel (b)), computed at the Courant Condition.

For historical reasons, the Yee solver \cite{Yee66numericalsolution} is one of the most used solver in our community and is based on a staggered second-order discretization of Maxwell's equation in space and time. As shown on panel (a), this solver induces a strong dispersion around axis $x$ and $z$ (i.e for propagation angles $\theta=0$ degree and $\theta=90$ degrees). \\
Several decades after Yee, Cole and Karkkainen proposed a solver\cite{Cole1997,Cole2002,Karkkainen2006}, which is  similar to the Yee's scheme but for which the spatial derivative of the Maxwell-Faraday equation is discretized with wider stencil, introducing a variable parameter $\beta$ in the expression of the finite difference in space (cf. Table \ref{Table_n}). A more general prescription for setting the parameter $\beta$ is given by Cowan in \cite{Cowan201161}.

The CKC solver is the particular case, where parameters $\Delta x=\Delta z$ and $\beta=0.25$. This particular solver is interesting because the Courant condition is equal to 1 and it allows $c\Delta t=\Delta x$ (=$\Delta z)$. Moreover, it is dispersion-free along $x$ and $z$. 

In principle, one may think that it is always possible to find a value of $\beta$ able to make the CKC solver non-dispersive in a direction of choice. However, we show in appendix that a such solver is unstable most of time. 

\subsection{Pseudo-spectral solvers}

Higher order-$p$ solvers \cite{Vincenti2016147} and their infinite limit order $p\rightarrow \infty$ pseudo-spectral solvers PSTD (Pseudo-Spectral Time Domain) \cite{Liu1997} can be used to suppress the dependency of the dispersion relation on the propagation angle of electromagnetic waves $\theta$. Indeed, when $p\rightarrow \infty$, the spatial discretized derivatives tend to the analytical solution in space, which completely suppresses the dependence of the refractive index with the angle  $\theta$.

The two different kinds of pseudo-spectral solvers that will be considered in this study are introduced below: 
\begin{enumerate}
\item \textbf{The PSTD solver} \cite{Liu1997}: The staggered Pseudo-Spectral Time Domain (PSTD) solver is the limit of the order-$p$ Yee solver when $p\rightarrow \infty$. In this scheme the solver is spectral in space but the time solve is still performed using second order leapfrog finite difference (cf. Table \ref{Table_n}). As a consequence, even though the PSTD solver has an isotropic dispersion relation, it is still dispersive and purely supraluminic ($n < 1$, $v_\varphi > c$). It also imposes a stringent Courant condition that can further increase computation time (see Table \ref{Table_n}). 
\item \textbf{The PSATD solver}  \cite{PSATD}: When solving Maxwell's equations in ($\textbf{k}$,t) space, it is possible to analytically integrate these equations in time and get an analytical solution for the spatio-temporal evolution of electric and magnetic fields. This analytical integration only assumes that electromagnetic sources are constant during one time step $\Delta t$, which is the basic assumption in PIC codes. This solver is called Pseudo-Spectral Analytical Time Domain (PSATD) solver. As opposed to the PSTD solver,  it is dispersion-free and imposes no courant condition in vacuum.  Note that when $\Delta t\rightarrow 0$ for a fixed spatial resolution, the PSTD solution converges to the PSATD one and the PSTD solver also becomes dispersion-free.
\end{enumerate}

In practice and as explained in section II. $3$ of this paper, it is better to use finite very large order-$p$ stencils in space instead of infinite order stencils for the approximation of the spatial derivative \cite{Vincenti2016147}. This significantly reduces truncation errors coming from our parallelization technique while still giving pseudo-spectral order $p\rightarrow \infty$ precision on almost the whole frequency spectrum. In ($\textbf{k}$,t) space, this simply consists in using a modified wavevector $\textbf{k}_p= \sum_{l=1}^{p/2} C_l^p \frac{sin\left(lk\Delta x\right)}{j\Delta x/2}$, where $C_l^p$ are the  Fonberg coefficients defined in \cite{Fornberg1990}, instead of $\textbf{k}$ in the PSATD and PSTD solvers so that $\textbf{k}_p=\hat{\nabla}_p$ is the Fourier transform of the order-p stencil $\nabla_p$ used in FDTD solvers (cf. Table \ref{Table_n}). In the following, we call PSATD $p$-order and PSTD p-order the PSATD and PSTD solvers with this modified spatial derivative $\textbf{k}_p$ in Fourier space. Note that the PSTD order-p solver in Fourier space is equivalent to the FDTD order-p solver in real space and for this reason has not been detailed in Table \ref{Table_n}. 

On Fig.\ref{optical_index}  (c) and (d), the refractive indices for the $128$-order PSTD and PSATD solvers are plotted. They are equivalent to infinite order PSATD and PSTD solvers on almost all the frequency domain  except near the Nyquist frequency (i.e $\textbf{k}\Delta x =\pi$). However, in practice the spatial resolution $\Delta x$ is chosen so that the harmonic spectrum is located in the lower half $\textbf{k}\Delta x <\pi/2$ of the frequency domain. 

\section{A model for harmonic angular deviation: a refraction model}
\begin{figure*}
\includegraphics[width=\linewidth]{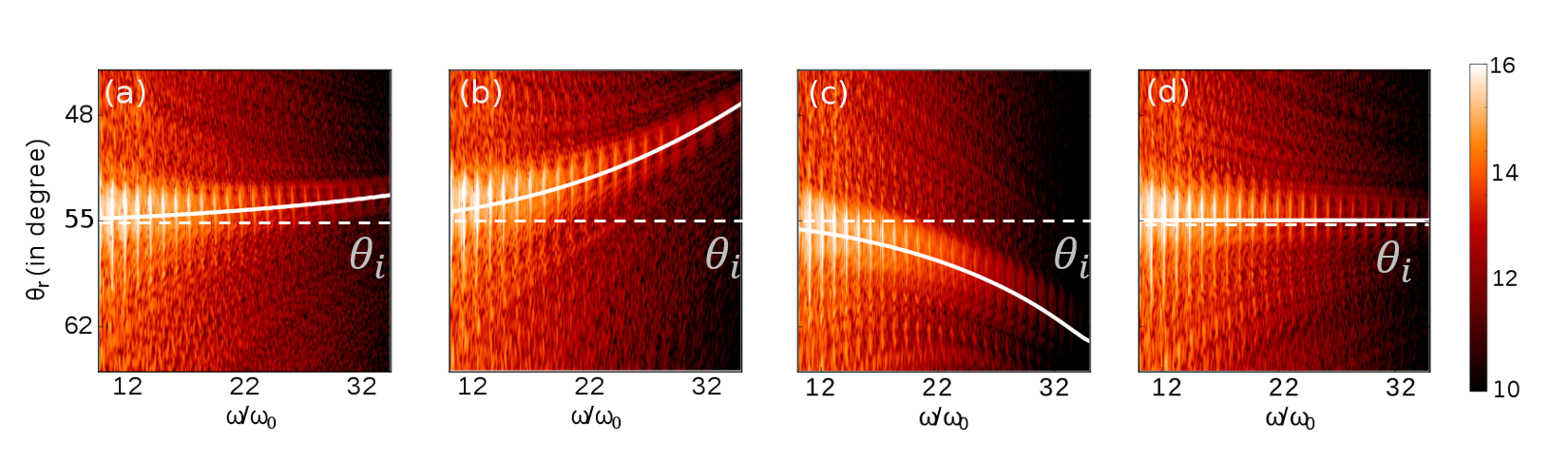} 
\caption{Angularly resolved spectra (log-scale) for different Maxwell's solvers in the range of $\omega_n/\omega_0=10$ to 35 for a spatial resolution of $\lambda_0/70$ and a time resolution fixed by the Courant condition. Panel (a): Yee solver. Panel (b): CKC solver. Panel (c): PSTD solver. Panel (d): PSATD solver. The laser incidence is $\theta_i=55$ degrees. The white solid line is the curve expected by the refraction model. For (a) and (b), $n_r >1$ and the harmonics get closer to the normal at higher frequency with a positive angle of refraction (cf Fig.\ref{target}). For (c), the PSTD solver numerical index $n_r<1$ and the refraction angle is negative. The panel (d) shows a dispersion-free spectrum obtained with the PSATD solver.}
\label{model_spectrum}
\end{figure*}

In this section, we investigate the effect of the numerical dispersion of the Maxwell solvers presented in the previous section on Doppler high harmonic spectra.  

\subsection{Effect of numerical dispersion on high harmonic generation}

Here, we present a simple toy-model that can be used to predict the angular deviation of high-order harmonics based on the Maxwell solver used and the spatio-temporal resolution of the simulation. This will allow us to further compute the best resolution that is needed to avoid angular dispersion effects for a given harmonic range. Our model is based on Snell-Descartes laws and its principle is sketched on Fig.\ref{target}. 

In the experiments, Doppler harmonics are all generated at the laser-plasma interface and enter vacuum with an angle $\theta_i=\theta_0$ equal to the angle of incidence of the laser on the target. As harmonics are generated exactly at the plasma-vacuum interface with the same angle $\theta_i$, one could see this situation as if each harmonic beam initially came from an "imaginary" non-dispersive medium of index $n_i=1$ and enter vacuum (see Fig.\ref{target}) with angle of incidence $\theta_i$.  As "real vacuum" is a non dispersive medium of refractive index $n_r=1$ for all frequencies, high-order harmonics will thus not be refracted and propagate in vacuum with the same angle $\theta_r=\theta_i $ (see panel (a)). 

By contrast, in the PIC simulation, Doppler harmonics are all generated at the laser-plasma interface but this time enter a dispersive medium ("numerical vacuum"), which has a refractive index $n_r[\omega, \theta_r]$ that depends on the frequency $\omega$ and on the propagation angle in the numerical medium $\theta_r$. As a consequence, different harmonic orders will be refracted by the plasma-vacuum interface at different angles $\theta_r$ (see panel (b)).The Snell's law presented below can be used to model this effect :
\begin{equation} 
\label{Snell}
n_i sin(\theta_i) = n_r[\omega, \theta_r] sin(\theta_r[\omega])
\end{equation}

where $n_i=1$ and $\theta_i=\theta_0$. Notice that in this particular case, $\theta_r$ also depends on $\omega$ due to the dispersion relation of the Maxwell solver. The unknown in equation (\ref{Snell}) to be determined for each harmonic frequency $\omega$ is thus $\theta_r[\omega]$. The medium refractive index $n_r$ is obtained from the numerical dispersion of Maxwell solvers and pictured on Fig.\ref{optical_index}. \\

With an iterative algorithm (e.g Newton-Raphson), we can easily find the numerical value of $\theta_r$ for every frequency $\omega$. In the following, we will compare deviations given by our model to results from PIC simulations for the different Maxwell solvers introduced in section III.

\subsection{Validation for different Maxwell solvers}

Fig. \ref{model_spectrum} shows angularly resolved harmonic spectra (color scale) obtained from PIC simulations for different Maxwell solvers (panels (a) to (d)). For each Maxwell solver, we superimposed on Fig. \ref{model_spectrum}  the angular deviation computed by solving equation (\ref{Snell}) for each frequency (white line) to the angularly resolved harmonic spectra.  For each Maxwell solver, our predictions agree very well with the angular deviation observed in simulation. 

\subsubsection{FDTD solvers}

For FDTD solvers such as Yee and CKC, the refractive index is always greater than $1$ (cf. Fig.\ref{optical_index} (a) and (b)). As a consequence and as the refractive index increases with frequency, the highest frequencies are deflected closer to the normal to the plasma mirror surface (cf. Fig. \ref{model_spectrum} (a) and (b)) as predicted by the Snell-Descartes law. 

\subsubsection{Pseudo-spectral solvers}

As opposed to FDTD solvers, the refractive index of vacuum for the PSTD-p solver is lower than 1 on almost all the frequency domain (cf. Fig.\ref{optical_index} (c)). As a consequence and because the refractive index decreases  with frequency, the highest frequencies are deflected further from the normal to the plasma mirror surface. Moreover, when light travels from a medium with a higher refractive index to one with a lower refractive index, Snell's law indicates that for an angle of incidence greater than a certain limit angle $\theta_l$, the wave should not pass through the interface and would be totally reflected. 
The reflective limit angle is obtained when $n_i sin(\theta_i) = n_r$. In practice, the total reflection appears at grazing incidence. For instance, the reflective limit angle, obtained for $\mathbf{k}\Delta x=\pi/2$, which corresponds to a refractive index $n_r^{\pi/2}=0.978$, is :
\begin{equation}
\theta_{l}^{\pi/2}=arcsin(n_r^{\pi/2})=77.96 \text{ deg}
\end{equation}

\begin{figure}[h!]
\includegraphics[width=\linewidth]{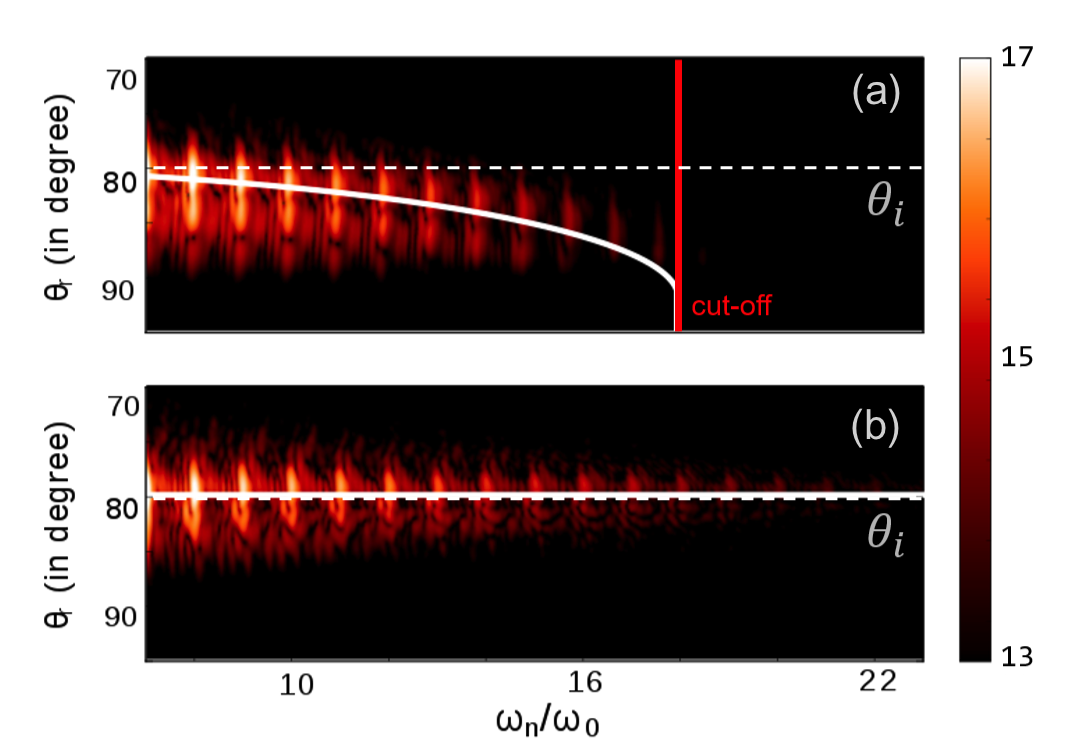} 
\caption{Angularly resolved spectrum (log-scale) for PSTD (panel (a)) and PSATD (panel (b)). The white solid line is the curve expected by the refraction model. For $\omega_n/\omega_0=17$, a numerical cut-off appears for PSTD due to the total reflection  on the interface. This cut-off is not physical and is absent with PSATD in the same configuration.}
\label{PSTD_refrac}
\end{figure}

Fig.\ref{PSTD_refrac}  shows the angularly resolved spectrum with a PIC simulation performed in same conditions as Fig. \ref{model_spectrum} but for a larger angle of incidence $\theta_i=80$ degrees. In this configuration and for order-128 PSTD solver, panel (a) illustrates that the highest harmonics are simply not generated and as predicted by our refraction model we observe a numerical spectrum cut-off around $\omega_n/\omega_0=17$. In contrast, for almost dispersion-free order 128-PSATD solver, panel (b) shows that harmonics above order 21$^{th}$ are generated. 

\section{Benefits of pseudo-spectral solvers}

For dispersive Maxwell solvers such as PSTD and FDTD, the model developed in section III can be extrapolated to calculate the spatio-temporal resolution required for mitigating the angular deviation $\Delta \theta$ in a given frequency range. In practice, we estimate that the angular deviation is negligible provided that the deviation $\Delta \theta<\theta_n/100$ is lower than a few percents of the total harmonic beam divergence $\theta_n$,  which is typically of the order of tens of mrads for a $100 mrads$ laser. This criterion imposes a maximal allowable deviation of around $5 mrad$ in these conditions. In the following, we will use this limit for computing the minimal resolution required with our model to avoid angular deviation effects. 

Along this line,  Table \ref{performance} presents the spatio-temporal resolutions  (and the corresponding memory/computation time) required for different Maxwell solvers in order to eliminate numerical dispersion effects until the 35$^{th}$ harmonic order for a typical 3D PIC simulations of Doppler harmonic generation (parameters detailed in caption of table \ref{performance}). Our model is applicable in 3D, based on the 3D phase velocity diagram of each solver. Switching from 2D to 3D is legit because the interaction is taking place in the ($x$,$z$) plane. 

In the Table, the memory and computation time required for dispersive FDTD and PSTD solvers are compared to the ones needed for an order-128 PSATD solver. The following performances are obtained with the same degree of optimization  The order-128 PSATD solver already gives converged numerical results for a spatial resolution of $\Delta x= \Delta z= \lambda_0/140$ as shown in appendix. On the MIRA supercomputer, the order-128 PSATD case at this resolution $r$ would require $T^{PSATD}_{r} = 2.5e8$ CPU hours and $M^{PSATD}_{r}=150$TBs of memory. For the same resolution , FDTD solver would perform $1.8$ times faster in 3D: $T^{FDTD}_{r}=T^{PSATD}_{r}/\xi$ with $\xi=1.8$. However, for FDTD solvers, our model shows that the resolution needed $r^{'}=A r$ to avoid angular dispersion would be $\times A$ higher than $r$, with $A>2$ (cf. Table \ref{performance}). Below, we estimate the time $T^{FDTD}_{r'}$ and memory $M^{FDTD}_{r'}$ required by the PIC algorithm in 3D with CKC and Yee solver when the spatial resolution $r$ is increased by a factor of $A$ to $r^{'}=Ar$:
\begin{itemize}
\item \textbf{The total memory} depends on the size of grid arrays (electromagnetic fields and sources) and the size of particle arrays (positions, velocities, weight). If $N_x$, $N_y$ and $N_z$ are the number of grid points for each axis and $N_p$, the number of particles, the space complexity of the PIC algorithm varies as $O(\alpha N_p+ \beta N_x N_y N_z )$  where $\alpha$ and $\beta$ are constant that depends on the number of grid arrays and particle quantities. For a constant number of particles per cell, $N_p$ can be expressed as $\gamma  N_x N_y N_z$ where $\gamma$ is the number of particles per cell. When the spatial resolution $r$ is increased by a factor $A$ on each axis, the memory $M^{FDTD}_{r'}$ is thus $A^3$ times greater than the initial one $M^{FDTD}_{r}$.  According to Table \ref{performance}, the resolution has to be $A=2.3$ times higher for FDTD solvers than pseudo-spectral solvers (PSTD and PSATD) in order to avoid angular deviation effects. The memory needed $M^{FDTD}_{r'}$  will thus be $2.3^3=12.1$ times higher (i.e almost $2$ PB of data) for FDTD solvers than pseudo-spectral solvers. This value (pictured in bold Table \ref{performance}) exceeds the entire available memory of the MIRA supercomputer at Argonne National Laboratory ($\simeq 800$ TB).
\end{itemize}
\begin{itemize}
\item \textbf{The time}. The time complexity of one time iteration of the PIC algorithm varies as $c_1=O(\delta N_x N_y N_z+\epsilon N_p)$ where $N_p=\gamma  N_x N_y N_z$, $\delta$ is a constant that depends on the number of instructions to be done per grid point and $\epsilon $ is a constant that depends on the number of operations to be done per particle (typically depends on the particle shape and deposition/gathering algorithm).  With FDTD solvers, the temporal resolution is fixed by the Courant condition and decreases when the spatial resolution decreases. As a consequence, if we increase the spatial resolution by A, the total time complexity of the simulation will be increased by $A^4$ and $T^{FDTD}_{r\prime}=A^4T^{FDTD}_{r}=A^4/\xi T^{PSATD}_{r}$. The ratios $T^{FDTD}_{r\prime}/T^{PSATD}_{r}$ are given for each FDTD solvers CKC and Yee (cf. Table \ref{performance}). Note that as the PSTD solver tends to the PSATD solver when the time step $\Delta t\rightarrow 0$, the spatial resolution needed is the same as PSATD but the time step has to be lower to achieve no angular deviation.  
\end{itemize}

\begin{table*}
\begin{tabular}{|c||c||c|c|c|}
\hline 
$\theta_i = 55$ degrees, $30-35^{th} order$ & PSATD & CKC & Yee & PSTD \\ 
\hline \hline 
\begin{tabular}{c} Resolution\\ $[$Factor$]$ \end{tabular} & $\lambda_0/140$ & \begin{tabular}{c} $\lambda_0/322$ \\  $[\times 2.3]$\end{tabular}& \begin{tabular}{c} $\lambda_0/322$ \\ $[\times 2.3]$  \end{tabular}& \begin{tabular}{c} $\lambda_0/140$ \\$[\times 1.0]$ \end{tabular}\\ 
\hline 
\begin{tabular}{c} Memory factor \\ $[$estimation (in TB) $]$ \end{tabular} & $[150TB]$ & \begin{tabular}{c} $\mathbf{\times 12}$  \\ $[1800TB]$  \end{tabular}& \begin{tabular}{c} $\mathbf{\times 12}$\\ $[1800TB]$  \end{tabular}& \begin{tabular}{c} $\times 1$ \\$[150TB]$ \end{tabular}\\ 
\hline 
\begin{tabular}{c} Time Factor\\ $[$estimation (in $M=10^6$ core hours)$]$ \end{tabular} & $[250M]$ & \begin{tabular}{c} $\mathbf{\times 16}$  \\ $\mathbf{[4000M]}$ \end{tabular}& \begin{tabular}{c} $\mathbf{\times 28}$ \\ $\mathbf{[7000M]}$   \end{tabular}& \begin{tabular}{c}$\times 9$ \\$[2250M]$   \end{tabular}\\ 
\hline
\end{tabular}  
\caption{Resolution, Memory and Time to solution needed to run a dispersion-free 3D simulation for different Maxwell solvers with a PIC code able to scale perfectly until a million cores.  These parameters are estimated for a typical 3D PIC simulation of Doppler harmonic generation, where the laser has an angle of incidence $\theta_i=55$ degrees on the plasma mirror, for which the harmonic generation efficiency is maximized. Besides, we assumed $10$ particles per cell, order $3$ current deposition/field gathering (QSP particle shapes). The simulation box dimensions of \{$L_x=40 \lambda$, $L_y=40 \lambda$, $L_z=60 \lambda$\}. The computation time and memory for the order-128 PSATD solver (left column) were estimated on the MIRA supercomputer and are taken as a reference case, for which we observed numerical convergence of harmonic spectra until harmonic orders $30-35$ (i.e. a deviation lower than $5$mrad). For other solvers, the figures in brackets are the ratio between the resolution (space/memory/time) required for the these solvers compared to the reference PSATD solver case. For PSTD, which converges to the PSATD solution when the time step $\Delta t\rightarrow 0$, we kept the same spatial resolution but we decreased the time step to $0.093\Delta x$.}
\label{performance}
\end{table*}

In short, Table \ref{performance} shows that FDTD solvers would require at maximum $\times12/\times28$ more memory/computation time than the PSATD solver and $\times 12/\times 3$ more memory/computation time than the PSTD solver. To avoid numerical dispersion, the PSTD time step needs to be significantly decreased until $0.110\Delta x$. Due to the absence of Courant condition, the PSATD solver then performs $\times 9$ better than the PSTD solvers and appears as the method of choice among pseudo-spectral solvers for modeling Doppler harmonic  on plasma mirrors. In the light of these figures, realistic 3D simulation of Doppler harmonic generation cannot be performed with a FDTD standard solver and pseudo-spectral solvers are a far better option, the PSATD solver offering the highest efficiency. \\

Notice that these speed-ups are lower estimates and would in principle even be higher as our model only gives spurious angular deviation but does not include other numerical effects induced by numerical dispersion such as group velocity dispersion. Based on these results and thanks to our massively parallel implementation of pseudo-spectral solvers, realistic simulations of $HHG$ are now becoming realistic on the largest machines available at the time of writing in terms of memory and computing power \cite{	00}. \\

Notice also that, in principle, the FDTD solver performances could be improved by using other target configurations, which suit better to each scheme. In the best case, the reflective wave must propagate along the best axis of the scheme. For example, in 3D, the Yee method is dispersion-free along the cube diagonal (\textit{i.e.} $k_x=k_y=k_z$). If the laser injection and the target design are carefully chosen to induce a reflection along one of the diagonal, a resolution of $\lambda/140$ would be enough to reduce numerical dispersion effects on the harmonic beams. However, for highly diverging harmonic beams or interaction configurations involving the emission of harmonic beams at many different angles \cite{PhysRevLett.108.113904, PhysRevLett.92.089901}, numerical dispersion could still severely affect harmonic beam properties. In addition, another practical problem is to redesign all the simulation and the diagnostics every time the angle of incidence or the Maxwell solver is changed which can be challenging for parametric studies. A potential solution would be to find a general formulation of FDTD solvers, which has a controllable parameter allowing to choose the dispersion-free axis of the solver. Unfortunately, we demonstrate in appendix that such solver cannot exist at second order and requires increasing the solver order, which reinforces the pertinence of high order solvers presented in this paper.  

\section{Conclusion}
\label{Conclusion}

In summary, standard FDTD Maxwell solvers are not suitable at reasonable computational cost to accurately describe Doppler harmonic generation with Particle-In-Cell codes, because they introduce numerical dispersion, causing angular deviation that significantly affects the harmonic spectrum. We showed that this angular deviation can be understood as a simple refraction of harmonic beams when these enter vacuum at the plasma-vacuum interface.  A simple model based on Snell-Descartes law was derived that can accurately reproduce the angular deviation observed in PIC simulations. This model can now be used to estimate the minimum resolution required to avoid this spurious deviation. The results of our model show that the required computing resources in 3D for these standard solvers exceed by far the ones of current petascale and future exascale supercomputer capacities.  In that case, a solution is to use dispersion-free pseudo-spectral solvers (as PSATD), for which there is no angular deviation even at moderate resolution. 

 \section*{Acknowledgement}

We thank Dr. Fabien Quere and Pr. Guy Bonnaud for fruitful discussions. This work was supported by the European Commission through the Marie Sk\l owdoska-Curie actions (Marie Curie IOF fellowship PICSSAR grant number 624543) as well as by the Director, Office of Science, Office of High Energy Physics, U.S. Dept. of Energy under Contract No. DE-AC02-05CH11231, the US-DOE SciDAC program ComPASS, and the US-DOE program CAMPA. An award of computer time (PICSSAR\_INCITE) was provided by the Innovative and Novel Computational Impact on Theory and Experiment (INCITE) program. This research used resources of the Argonne Leadership Computing Facility, which is a DOE Office of Science User Facility supported under Contract DE-AC02-06CH11357 and of the National Energy Research Scientific Computing Center, a DOE Office of Science User Facility supported by the Office of Science of the U.S. Department of Energy under Contract No. DE-AC02-05CH11231.

This document was prepared as an account of work sponsored in part
by the United States Government. While this document is believed to
contain correct information, neither the United States Government
nor any agency thereof, nor The Regents of the University of California,
nor any of their employees, nor the authors makes any warranty, express
or implied, or assumes any legal responsibility for the accuracy,
completeness, or usefulness of any information, apparatus, product,
or process disclosed, or represents that its use would not infringe
privately owned rights. Reference herein to any specific commercial
product, process, or service by its trade name, trademark, manufacturer,
or otherwise, does not necessarily constitute or imply its endorsement,
recommendation, or favoring by the United States Government or any
agency thereof, or The Regents of the University of California. The
views and opinions of authors expressed herein do not necessarily
state or reflect those of the United States Government or any agency
thereof or The Regents of the University of California.

\bibliography{biblio}

\begin{appendix}
\section{Dispersion-free FDTD solver for a given angle}

It has been suggested that a dispersion-free FDTD solver for a given angle exists and is stable at the Courant condition. \\
In the upcoming section, we show that it is not possible to find in 2D a set of coefficients to reach a stable second-order FDTD scheme depending on the angle $\theta=\arctan (k_z/k_x)$, which presents no dispersion for this angle.\\
To do so, we are using the Cole-Karkkainen solver with a $\beta^\prime$ defined in \cite{Vay20115908} and $\Delta x=\Delta z$. Note that the $\beta^\prime$ is equal to $-4 \: \beta$ and so the CKC solver presents a $\beta^\prime=-1$. With $\eta=\Delta x/ c \Delta t$, the dispersion equation is :

\begin{equation}\label{eq:dispersion}
\begin{aligned}
sin^2 \left( \frac{\eta \tilde{k}}{2} \right) = & \eta^2 \left[ sin^2 \left( \frac{ \tilde{k} cos(\theta)}{2}\right) + sin^2 \left(\frac{ \tilde{k} sin(\theta)}{2} \right) \right. \\ 
& \left. + \beta^\prime sin^2 \left( \frac{ \tilde{k} cos(\theta)}{2}\right) \; sin^2 \left(\frac{ \tilde{k} sin(\theta)}{2} \right) \right]
\end{aligned}
\end{equation}

\noindent  where $\tilde{k}$ is defined as 
\begin{equation}
  \left\{
      \begin{aligned}
&k_x \Delta x=\tilde{k} cos(\theta)\\ 
&k_z \Delta z=\tilde{k} sin(\theta)
      \end{aligned}
    \right.    
\end{equation}

The final solver must satisfy this equation for any $\tilde{k}$ and in particular when  $\tilde{k} \rightarrow 0$. This relation is verified for every $\eta$ and $\beta^\prime$ at the second order. To obtain a solver without any dispersion for the right angle,  we developp it to the $4^{th}$ and the $6^{th}$ order in $\tilde{k}$ near $0$. \\
After developpment, we find the following system :

\begin{equation}
  \left\{
      \begin{aligned}
\text{order 4 : }&\eta^2 = 1 - (2+3 \: \beta^\prime) \:  cos^2(\theta) \:  sin^2(\theta)\\ 
\text{order 6 : }&\eta^4 = 1 - (3+\frac{15}{2} \: \beta^\prime) \:  cos^2(\theta) \:  sin^2(\theta)
      \end{aligned}
    \right.    
\end{equation}

\begin{figure}[h!]
\includegraphics[width=\linewidth]{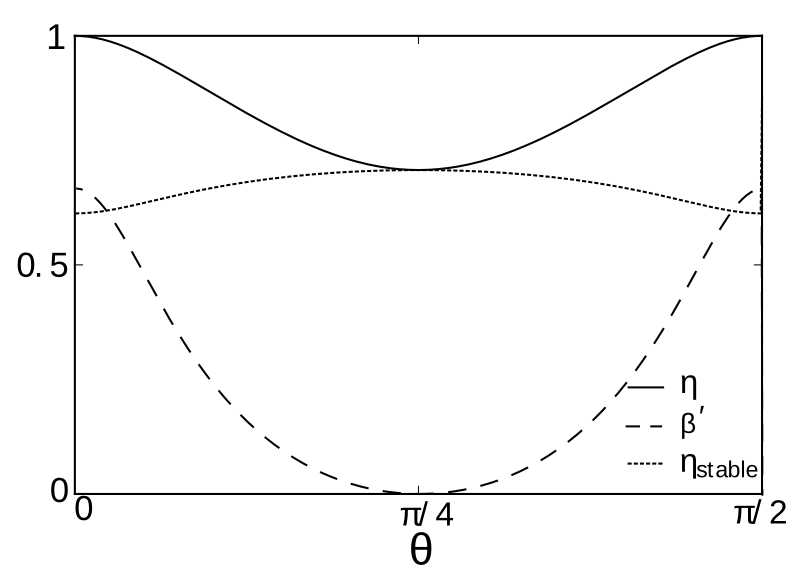} 
\caption{Variations of the coefficients solution of the system \ref{eq:coef_FDTD} in function of $\theta$, given incident angle. For a stable scheme, the full line curve must be below the dot line curve \textit{i.e.} $\eta \leq \eta_{stable}$. However, this condition is valid only for $\theta=\pi/4$ and corresponds to the Yee scheme. }
\label{fig:coef_FDTD}
\end{figure}

This system can be solved analytically for a given $\theta$ and we found :
\begin{equation}\label{eq:coef_FDTD}
  \left\{
      \begin{aligned}
&\eta = \frac{1}{2}\sqrt{5 - \sqrt{1 + 32 cos^2(\theta) \:  sin^2(\theta)}}\\ 
&\beta^\prime = \frac{1 - \eta^2 - 2 cos^2(\theta) \:  sin^2(\theta)}{3 cos^2(\theta) \:  sin^2(\theta)}
      \end{aligned}
    \right.    
\end{equation}

Note that for $\theta$ approaching $\pi/4$, we found the Yee scheme with $\eta=1/\sqrt{2}$ and $\beta^\prime=0$.\\

If this solution is stable for this particular $\eta$ then the final solver is given with this particular set of coefficients. \\
We make the ansatz that the most unstable modes propagate at the Nyquist wavelength where $k_x=k_z=\pi$ :
\begin{equation}
\eta_{stable}^2 \leq \frac{1}{2+\beta^\prime}
\end{equation}

The profile of the coefficients $\eta$, $\beta^\prime$ and $\eta_{stable}$ is plotted Fig. \ref{fig:coef_FDTD}. The condition of stability is defined such as $\eta \leq \eta_{stable}$. This condition is never verified, except for $\theta=\pi/4$, Yee scheme configuration. \\

According to Fig. \ref{fig:coef_FDTD}, $\beta^\prime$ is always positive and it seems to be inconsistent with the CKC solver, which presents $\beta^\prime=-1$. However, one might notice that $\beta^\prime$ is absent in the equation \ref{eq:dispersion}, when $\theta=0$ or $\pi/2$. $\beta^\prime$ may take every value and in particular $-1$, which ensures $\eta=\eta_{stable}=1$.\\

It does not exist a stable order 2 solver, depending on the incident angle $\theta$, that is perfectly accurate for this angle. A solution might be to increase the number of coefficients and then the order of the solver. When the order $p \rightarrow \infty$, the solver tends toward a pseudo-spectral solver. \\

\section{Convergence of PSATD-$128$ solver}

\begin{figure}[h!]
\centering
\includegraphics[width=\linewidth]{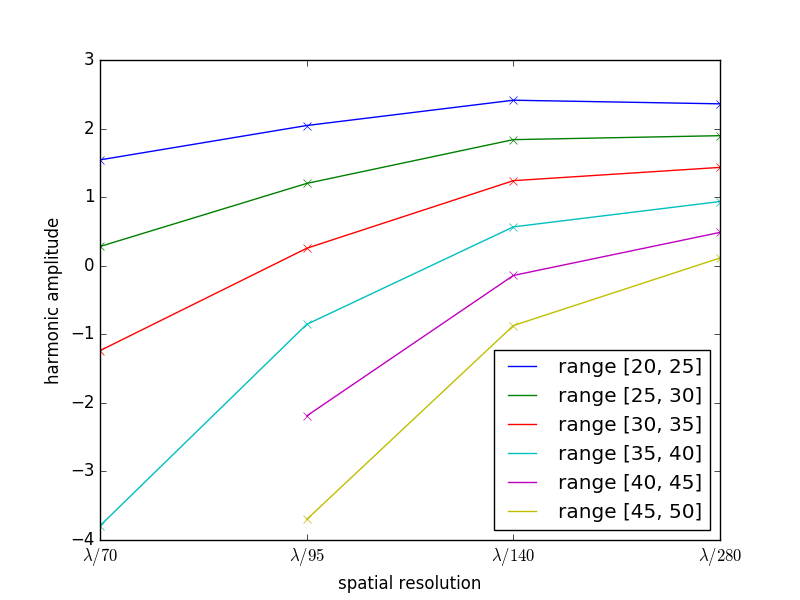}
\caption{\label{fig:convergence_cases} Evolution of high harmonic generation efficiency as a function of spatial resolution in the case of the PSATD solver. Numerical/physical parameters of the simulation are the same as in the manuscript. Harmonic generation efficiency for different harmonic range is obtained by integrating the angularly-resolved spectrum in angle. The $y$-axis is in in log scale. Until at least the $30-35^{th}$ harmonic, this figure shows convergence of the high harmonic spectrum at a resolution of $\lambda/140$ in $2$D. }
\end{figure}

We assessed convergence in $2$D based on (i) the suppression of spurious angular dispersion and (ii) the convergence of high harmonic generation efficiency for harmonic orders up to $30-35$. In Fig. \ref{fig:convergence_cases} above we show the evolution of high harmonic efficiency with resolution for different harmonic ranges in the case of the PSATD solver (order $128$) obtained from 2D PIC simulations with WARP+PXR. Fig. \ref{fig:convergence_cases} clearly shows convergence for harmonic orders ranging from $30$ to $35$ at a resolution of $\lambda/140$. This resolution is the one used to estimate speed-ups of PSATD vs other FDTD schemes in the manuscript. 

\end{appendix}

\end{document}